\documentclass[prl,twocolumn,superscriptaddress,preprintnumbers,amsmath,amssymb]{revtex4}
\usepackage{graphicx}
\usepackage{amsmath}
\usepackage{latexsym}
\usepackage{dcolumn}
\usepackage{bm}
\usepackage{float}
\usepackage{graphicx,here}
\usepackage{color}
\usepackage{hyperref}
\usepackage{bookmark}
\usepackage{color}
\hypersetup{
colorlinks=true,
citecolor=blue
}

\emergencystretch=\maxdimen
\begin{document}

\title{\bf Layer-dependent Raman spectroscopy of ultrathin Ta$_2$Pd$_3$Te$_5$}

\affiliation{Institute of Physics, Chinese Academy of Sciences, Beijing 100190, China}
\affiliation{Advanced Research Institute of Multidisciplinary Science, Beijing Institute of Technology, Beijing 100081, China}
\affiliation{School of Physical Sciences, University of Chinese Academy of Sciences, Beijing 100049, China}
\affiliation{Center of Materials Science and Optoelectronics Engineering, University of Chinese Academy of Sciences, Beijing 100049, China}
\affiliation{Songshan Lake Materials Laboratory, Dongguan, Guangdong 523808, China}
\affiliation{Interdisciplinary Institute of Light-Element Quantum Materials and Research Center for Light-Element Advanced Materials, Peking University, Beijing, 100871, China}

\author{Zhenyu Sun$^\dag$}
\affiliation{Institute of Physics, Chinese Academy of Sciences, Beijing 100190, China}
\affiliation{School of Physical Sciences, University of Chinese Academy of Sciences, Beijing 100049, China}
\author{Zhaopeng Guo$^\dag$}
\affiliation{Institute of Physics, Chinese Academy of Sciences, Beijing 100190, China}
\affiliation{School of Physical Sciences, University of Chinese Academy of Sciences, Beijing 100049, China}
\author{Dayu Yan}
\affiliation{Institute of Physics, Chinese Academy of Sciences, Beijing 100190, China}
\affiliation{School of Physical Sciences, University of Chinese Academy of Sciences, Beijing 100049, China}
\author{Peng Cheng}
\affiliation{Institute of Physics, Chinese Academy of Sciences, Beijing 100190, China}
\affiliation{School of Physical Sciences, University of Chinese Academy of Sciences, Beijing 100049, China}
\author{Lan Chen}
\affiliation{Institute of Physics, Chinese Academy of Sciences, Beijing 100190, China}
\affiliation{School of Physical Sciences, University of Chinese Academy of Sciences, Beijing 100049, China}
\affiliation{Songshan Lake Materials Laboratory, Dongguan, Guangdong 523808, China}
\author{Youguo Shi}
\affiliation{Institute of Physics, Chinese Academy of Sciences, Beijing 100190, China}
\affiliation{School of Physical Sciences, University of Chinese Academy of Sciences, Beijing 100049, China}
\affiliation{Center of Materials Science and Optoelectronics Engineering, University of Chinese Academy of Sciences, Beijing 100049, China}
\affiliation{Songshan Lake Materials Laboratory, Dongguan, Guangdong 523808, China}
\author{Yuan Huang}\thanks{yhuang@bit.edu.cn}
\affiliation{Advanced Research Institute of Multidisciplinary Science, Beijing Institute of Technology, Beijing 100081, China}
\author{Zhijun Wang}
\affiliation{Institute of Physics, Chinese Academy of Sciences, Beijing 100190, China}
\affiliation{School of Physical Sciences, University of Chinese Academy of Sciences, Beijing 100049, China}
\author{Kehui Wu}\thanks{khwu@iphy.ac.cn}
\affiliation{Institute of Physics, Chinese Academy of Sciences, Beijing 100190, China}
\affiliation{School of Physical Sciences, University of Chinese Academy of Sciences, Beijing 100049, China}
\affiliation{Songshan Lake Materials Laboratory, Dongguan, Guangdong 523808, China}
\affiliation{Interdisciplinary Institute of Light-Element Quantum Materials and Research Center for Light-Element Advanced Materials, Peking University, Beijing, 100871, China}
\author{Baojie Feng}\thanks{bjfeng@iphy.ac.cn}
\affiliation{Institute of Physics, Chinese Academy of Sciences, Beijing 100190, China}
\affiliation{School of Physical Sciences, University of Chinese Academy of Sciences, Beijing 100049, China}
\affiliation{Interdisciplinary Institute of Light-Element Quantum Materials and Research Center for Light-Element Advanced Materials, Peking University, Beijing, 100871, China}

\date{\today}

\clearpage

\begin{abstract}
Two-dimensional topological insulators (2DTIs) or quantum spin Hall insulators are attracting increasing attention due to their potential applications in next-generation spintronic devices. Despite their promising prospects, realizable 2DTIs are still limited. Recently, Ta$_2$Pd$_3$Te$_5$, a semiconducting van der Waals material, has shown spectroscopic evidence of quantum spin Hall states. However, achieving controlled preparation of few- to monolayer samples, a crucial step in realizing quantum spin Hall devices, has not yet been achieved. In this work, we fabricated few- to monolayer Ta$_2$Pd$_3$Te$_5$ and performed systematic thickness- and temperature-dependent Raman spectroscopy measurements. Our results demonstrate that Raman spectra can provide valuable information to determine the thickness of Ta$_2$Pd$_3$Te$_5$ thin flakes. Moreover, our angle-resolved polarized Raman (ARPR) spectroscopy measurements show that the intensities of the Raman peaks are strongly anisotropic due to the quasi-one-dimensional atomic structure, providing a straightforward method to determine its crystalline orientation. Our findings may stimulate further efforts to realize quantum devices based on few or monolayer Ta$_2$Pd$_3$Te$_5$.
\end{abstract}

\maketitle

\section{Introduction}

Topological quantum materials (TQMs) have been a topic of major interest in condensed matter physics over the last decade~\cite{QiXL2011,WangJ2017,HasanMZ2010}. These materials possess topologically nontrivial band structures, which can give rise to a variety of physical phenomena, such as high carrier mobility, quantum spin/anomalous Hall effects, and axion insulating states. Two-dimensional topological insulators (2DTIs) are a special class of TQMs with insulating bulk states and topological edge states. While the first type of 2DTI, quantum well systems such as HgTe/CdTe~\cite{KonigM2007} and InAs/GaSb~\cite{SpantonEM2014,DuL2015}, posed challenges due to the stringent requirements for composition and thickness control, recent spectroscopic evidence suggests that several graphene-like 2D materials, such as stanene~\cite{ZhuF2015,DengJ2018} and bismuthene~\cite{ReisF2017}, can be synthesized in ultrahigh vacuum by molecular beam epitaxy. However, these epitaxial 2D materials typically interact strongly with substrates and are challenging to transfer to insulating substrates for device fabrication and transport measurements.

In recent years, van der Waals materials represented by 1T$^\prime$-MX$_2$ (M=Mo, W; X=S, Se, Te) have been predicted to be 2DTIs~\cite{QianX2014}. However, despite these predictions, only monolayer 1T$^\prime$-WTe$_2$ has provided transport evidence for 2DTIs~\cite{WuS2018}. In multilayer samples, QSH effects disappear due to the bulk band gap closing. Nevertheless, a new van der Waals material, Ta$_2$Pd$_3$Te$_5$, has been recently predicted to be a 2DTI~\cite{GuoZ2021,GuoZ2022}. Interestingly, both bulk and monolayer samples host an inverted gap, which makes it possible to fabricate QSH devices using multilayer samples. Compared to monolayer materials, multilayer materials have numerous advantages such as increased stability in ambient conditions and greater flexibility in thickness and twist angles. Angle-resolved photoemission spectroscopy (ARPES) and scanning tunneling spectroscopy (STS) measurements have confirmed the existence of a bulk band gap and topological edge states in Ta$_2$Pd$_3$Te$_5$~\cite{WangX2021}. Moreover, recent transport measurements have demonstrated the existence of Tomonoga-Luttinger liquid states in Ta$_2$Pd$_3$Te$_5$ devices, indicating the coexistence of topological order and strong correlation effects~\cite{WangA2022}. However, experimental realization of few to monolayer Ta$_2$Pd$_3$Te$_5$ remains challenging, yet is crucial for fabricating Ta$_2$Pd$_3$Te$_5$-based quantum devices.

In this work, we successfully obtained large-scale few and monolayer Ta$_2$Pd$_3$Te$_5$ flakes using the Au-assisted mechanical exfoliation method. Our study employed a combination of Raman spectroscopy measurements and theoretical calculations to identify all active vibrational modes and demonstrate their strong anisotropy. Additionally, our investigation of the thickness-, temperature-, and polarization-dependent behavior of Raman peaks provided critical information for determining the thickness and crystalline orientation of Ta$_2$Pd$_3$Te$_5$ thin flakes.

\section{Methods}

Ta$_2$Pd$_3$Te$_5$ single crystals were synthesized using the self-flux method \cite{GuoZ2021}. High-purity Ta, Pd, and Te powders were mixed in an alumina crucible with a molar ratio of 2:4.5:7.5, then sealed in a quartz tube, and heated to 950 $^{\circ}$C over 10 hours, dwelt for 2 days, and slowly cooled down to 800 $^{\circ}$C. The extra flux was removed by centrifuging at 800 $^{\circ}$C. Large-area few to monolayer Ta$_2$Pd$_3$Te$_5$ flakes were obtained by the Au-assisted mechanical exfoliation technique \cite{HuangY2020}. The SiO$_2$/Si substrate was cleaned with oxygen plasma treatment. An ultrathin Ti layer was prepared on the SiO$_2$/Si substrate to improve the smoothness and adhesion of the Au film. Afterward, a tape with a freshly cleaved Ta$_2$Pd$_3$Te$_5$ surface was pressed onto the Au film. Atomically thin Ta$_2$Pd$_3$Te$_5$ flakes can be obtained after removing the tape.

Optical images were acquired with an Olympus MX50 optical microscope. Atomic force microscopy (AFM) measurements were performed using a commercial AFM system (Oxford, Asylum Research Cypher S). Scanning electron microscopy (SEM) measurements and energy-dispersive X-ray spectroscopy (EDS) mapping were conducted using a commercial field-emission SEM (Hitachi, SU5000). Raman spectroscopy measurements were performed using a micro-Raman system (Horiba LabRam HR Evolution) with two laser sources (532 and 633 nm). The illumination power was approximately 1 mW. The peak at 520 cm$^{-1}$ originating from the Si substrate was used to calibrate each spectrum.

First-principles calculations were performed using the Vienna ab initio simulation package (VASP)~\cite{KresseG1996-1,KresseG1996-2} within the projector-augmented wave (PAW) framework \cite{BlochlPE1994,KresseG1999}. The generalized gradient approximation (GGA) in Perdew-Burke-Ernzerhof (PBE) functional theory~\cite{PerdewJP1996} was used in the calculations. The energy cutoff of the plane wave basis was set to 500 eV. A $1\times4\times1$ supercell was built to calculate the Raman frequencies and irreducible representations using the PHONOPY package~\cite{TogoA2008}.

\section{Results and discussion}

Ta$_2$Pd$_3$Te$_5$ is a van der Waals material with a layered structure (space group: \emph{Pnma}). Its bulk unit cell consists of two monolayers, with each monolayer exhibiting a two-dimensional topological insulator behavior~\cite{GuoZ2021}. The Ta$_2$Pd$_3$Te$_5$ layers are stacked along the $[100]$ direction, and the adjacent monolayers are held together by weak van der Waals interactions with a binding energy of approximately 19.6 meV/\AA$^2$, similar to other van der Waals materials~\cite{GuoZ2021}. Hence, few and even monolayer Ta$_2$Pd$_3$Te$_5$ can be obtained through mechanical exfoliation for device fabrication. However, exfoliating large-scale ultrathin samples on a pristine SiO$_2$/Si substrate is challenging. Therefore, we employed an Au-assisted method that is effective in exfoliating transition metal dichalcogenides~\cite{DesaiSB2016,VelickyM2018,LiuF2020,HuangY2020} due to the enhanced adhesion between Te atoms in Ta$_2$Pd$_3$Te$_5$ and Au. Using this method, we obtained few and monolayer Ta$_2$Pd$_3$Te$_5$ with high yield. Fig. 1(a) shows a schematic drawing of the crystal structure of Ta$_2$Pd$_3$Te$_5$, while Fig. 1(b) displays a typical optical image of exfoliated Ta$_2$Pd$_3$Te$_5$ thin flakes on a Au/SiO$_2$/Si substrate, indicating different thicknesses (1L to 10L). The thickness-dependent optical reflectivity from the green channel image is presented in Fig. 1(c), and the linear relation is consistent with the Fresnel-law-based model~\cite{BlakeP2007}. The quality of the samples was confirmed by atomic force microscopy (AFM) and scanning electron microscopy (SEM) measurements, as shown in Fig. S2 and Fig. S3 within the Supplemental Material (SM)~\cite{SM}. Although we also exfoliated Ta$_2$Pd$_3$Te$_5$ directly on the SiO$_2$/Si substrate, the thinnest flakes we obtained were $\sim$4 nm thick, approximately 6 layers of monolayer Ta$_2$Pd$_3$Te$_5$.

\begin{figure}[t]
\centering
\includegraphics[width=8.5 cm]{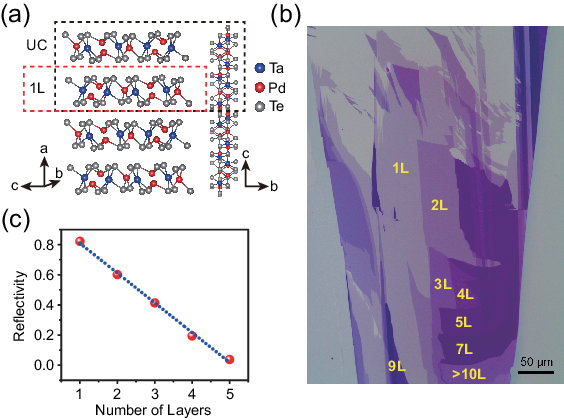}
\caption{(a) Schematic drawing of the crystal structure of Ta$_2$Pd$_3$Te$_5$. (b) Optical image of exfoliated Ta$_2$Pd$_3$Te$_5$ thin flakes on an Au/SiO$_2$/Si substrate, with the number of layers indicated. (c) Relative optical reflectivity of 1-5L extracted from the green channel image. The blue dotted line is a linear fit of the optical reflectivity versus the number of layers.}
\end{figure}

To investigate the vibrational modes of Ta$_2$Pd$_3$Te$_5$, we first conducted Raman spectroscopy measurements on bulk materials. The results are presented in Fig. 2(a), which show several Raman peaks. The peak positions are found to be independent of the incident light wavelength, despite variations in their relative intensity, as shown in Fig. S4 within SM~\cite{SM}. To determine the vibrational modes of each Raman peak, we performed first-principles calculations, and the outcomes are summarized in Table S1 within SM~\cite{SM}. By comparing the experimental and theoretical results, we identified 9 A$_g$ modes and 3 B$_{3g}$ modes, as indicated in Fig. 2(a). Based on the symmetry of the D$_{2h}$ point group, the irreducible representation of the zone center phonon modes are $\Gamma=20A_g+10B_{1g}+20B_{1u}+20B_{2g}+10B_{2u}+10B_{3g}+20B_{3u}$, where the A$_g$, B$_{1g}$, B$_{2g}$, and B$_{3g}$ modes are Raman active~\cite{Loudon1964}. Due to the backscattering geometry in our measurements, only the A$_g$ and B$_{3g}$ modes can be detected according to the selection rules. These symmetry analyses are consistent with our calculation results. The atomic displacements of these vibrational modes are illustrated in Fig. 2(b) and Fig. S5 within SM~\cite{SM}, noting that all the B$_{3g}$ modes correspond to in-plane vibrations.

\begin{figure}[t]
\centering
\includegraphics[width=8.5 cm]{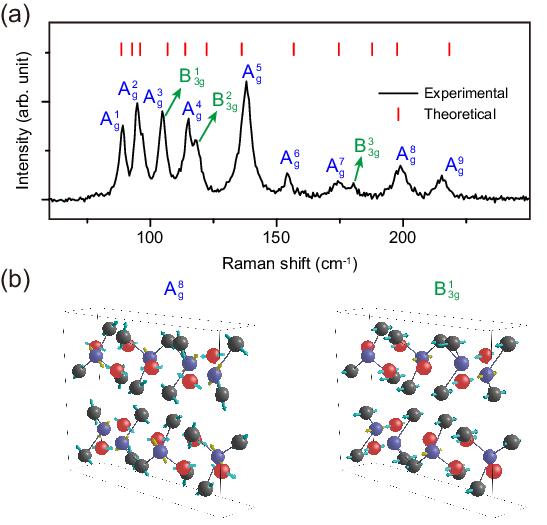}
\caption{(a) Typical Raman spectrum of bulk Ta$_2$Pd$_3$Te$_5$ measured with an excitation wavelength of 532 nm, with the calculated Raman peaks indicated by red lines. (b) Atomic displacements of the A$_g^8$ and B$_{3g}^1$ Raman modes.}
\end{figure}

After clarifying the vibrational modes of bulk Ta$_2$Pd$_3$Te$_5$, we proceeded to investigate ultrathin Ta$_2$Pd$_3$Te$_5$ flakes. Figure 3(a) displays the Raman spectra measured on different layers. As the thickness decreases, the intensity of the Raman peaks gradually increases and then decreases towards the monolayer limit. This phenomenon has also been observed in graphene and MoS$_2$ and can be attributed to the suppression of constructive interference enhancement due to the absorption of incident and scattered light in bulk materials~\cite{WangYY2008,YuanH2019}. Based on the thickness-dependent Raman spectra, we have several important observations. First, the Raman peak intensity of monolayer samples is significantly weaker than that of few-layer or bulk samples, which may originate from the strong charge transfer or chemical bonding at the Ta$_2$Pd$_3$Te$_5$/Au interface. The suppression of Raman signal in monolayer samples has also been reported in exfoliated 2D materials on Au film~\cite{LiG2022}. Second, most Raman peaks show no detectable shift with respect to thickness, except for the out-of-plane A$_g^6$ mode at $\sim$154 cm$^{-1}$, which exhibits a blue shift with increasing thickness, as shown in Fig. S6 within SM~\cite{SM}. The stiffening of the A$_g^6$ mode in thicker films may be caused by the increased interlayer interactions with additional layers~\cite{LeeC2010,YamamotoM2014}. Third, the intensity of the A$_g^1$ mode at 89 cm$^{-1}$ almost disappears when the thickness is reduced to 2L, as indicated by the black arrows in Fig. 3(a). We will demonstrate later that the A$_g^1$ mode reappears in 2L samples at low temperature, which can be utilized to distinguish 1L and 2L samples. Fourth, the Raman intensity ratio of Ta$_2$Pd$_3$Te$_5$ with respect to Si (520 cm$^{-1}$) linearly increases with increasing thickness, as shown in Fig. 3(b). This behavior is consistent with other van der Waals materials~\cite{HuangY2014}.

\begin{figure*}[htb]
\includegraphics[width=17 cm]{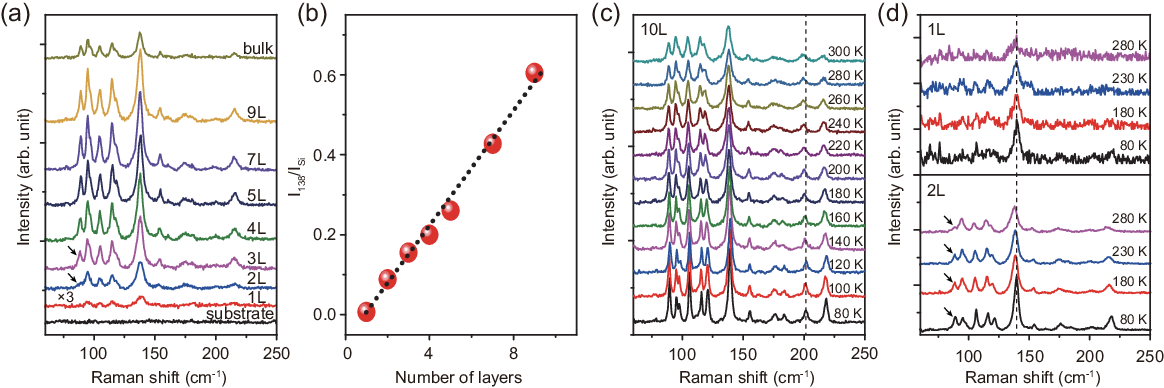}
\caption{(a) Raman spectra of 1-10L Ta$_2$Pd$_3$Te$_5$ with the excitation wavelength set to 532 nm and the polarization parallel to the $c$ axis of Ta$_2$Pd$_3$Te$_5$. (b) Raman intensity ratio of the peak at 138 cm$^{-1}$ with respect to Si as a function of the number of layers. (c) Temperature-dependent Raman spectra of the 10L sample in the range from 80 K to 300 K. (d) Raman spectra of the 1L and 2L samples measured at 80, 180, 230, and 280 K, respectively.}
\end{figure*}

We conducted a detailed investigation into the temperature dependence of the vibrational modes of Ta$_2$Pd$_3$Te$_5$ using Raman spectroscopy measurements from 80 K to 300 K, as presented in Fig. 3(c). The results reported here are for the 10L samples due to the greater Raman signal intensity. We observed a blue shift and a noticeable linewidth sharpening for all Raman peaks as the temperature decreased, as shown in Fig. S11 within SM~\cite{SM}. This behavior indicates the suppression of anharmonic effects during cooling~\cite{MenendezJ1984,SarkarS2020}. Our results suggest that there is no structural phase transition, which is in agreement with the electrical transport measurements of bulk Ta$_2$Pd$_3$Te$_5$~\cite{WangX2021}. However, a careful analysis for Raman peak position and linewidth~\cite{Sorb2020} indicates that there might be an isostructural transition at around 150 K (Fig. S10-S12 within SM~\cite{SM}). Since there is no obvious thickness-dependent shift for most Raman peaks, the results are representative and reliable for multi-layer or even bulk crystals. For the 1L sample, the A$_g^5$ mode had a negligible shift, which may be caused by the strong interaction of the Au substrate, as shown in the top panel of Fig. 3(d). Such interaction might be a covalent-like quasi-bonding between Te and Au atoms~\cite{HuangY2020}, which will suppress atomic vibrations at the interface. However, for the 2L sample, a blue shift with decreasing temperature was observed, which is analogous to thick samples, as shown in the bottom panel of Fig. 3(d) and Fig. S13 within SM~\cite{SM}. Notably, the intensity of the A$_g^1$ mode in 2L samples increased at low temperatures, as indicated by the black arrows in Fig. 3(d). Further theoretical and experimental efforts are required to understand the origin of this behavior.

We used the ARPR technique to investigate the symmetry of the vibrational modes of Ta$_2$Pd$_3$Te$_5$. The experimental setup for ARPR spectroscopy measurements is illustrated in Fig. 4(a). The samples were rotated continuously along the perpendicular direction (the $[100]$ axis), while the incident light was kept fixed. The scattered light was collected while the analyzer direction was set parallel or perpendicular to the incident direction. The crystal orientations of Ta$_2$Pd$_3$Te$_5$ were determined by Laue diffraction measurements, as shown in Fig. S14 within SM~\cite{SM}. The $[010]$ direction was defined as 0$^\circ$ in our measurements. The ARPR measurement results are summarized in Fig. 4(b) and 4(c). We observed that the intensity of most Raman peaks exhibited periodic oscillations with respect to the azimuth angle. In the parallel-polarized configuration, all the A$_g$ modes exhibited two-fold symmetry (Fig. S15 within SM~\cite{SM}). For example, the Raman intensity of the A$_g^8$ mode reached the maximum at 0$^\circ$ and 180$^\circ$, as shown in Fig. 4(d). In the cross-polarized configuration, these A$_g$ modes reached maximum intensity at 45$^\circ$, 135$^\circ$, 225$^\circ$ and 315$^\circ$ and formed a four-lobed shape in the polar graph, as shown in Fig. S16 within SM~\cite{SM} and Fig. 4(e). As for the B$_{3g}$ modes, we observed a four-fold symmetric angular-dependent evolution in both configurations, but the lobe was rotated by 45$^\circ$, as shown in Fig. 4(f). These results indicated that the vibrational modes of Ta$_2$Pd$_3$Te$_5$ are strongly anisotropic.

\begin{figure*}[htb]
\includegraphics[width=17 cm]{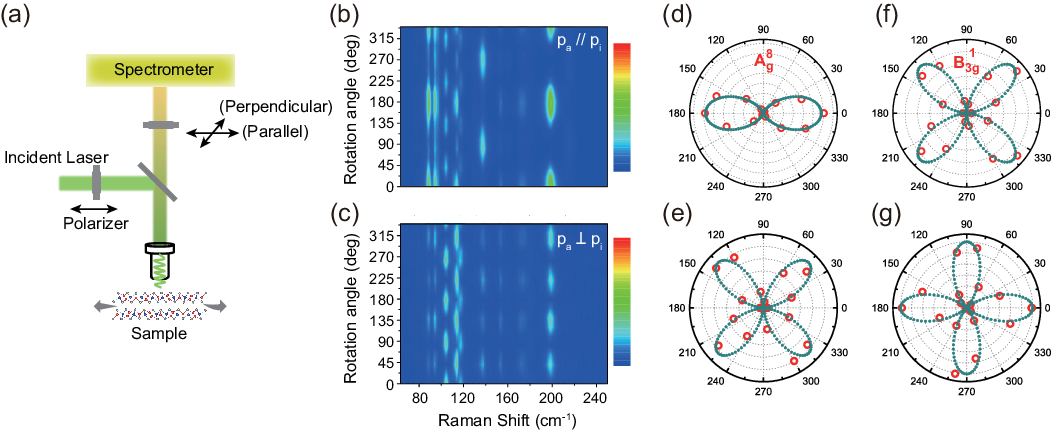}
\caption{(a) Schematic diagram of the ARPR measurement setup. (b) and (c) Angular dependence of Raman intensities on the 10L sample in the parallel-polarized (b) and cross-polarized (c) configuration, respectively. (d) and (e) Normalized Raman intensities (red circles) of the A$_g^8$ mode versus the azimuth angle in the parallel-polarized (d) and cross-polarized (e) configurations, respectively. (f) and (g) Normalized Raman intensities (red circles) of the B$_{3g}$ mode versus the azimuth angle in the parallel-polarized (f) and cross-polarized (g) configurations, respectively. The fitted results are indicated by green dotted lines in (d)-(g). The excitation wavelength is 532 nm.}
\end{figure*}

The results presented above can be well-explained using the classical Placzek approximation~\cite{Albrecht1961,Loudon1964}, in which the Raman intensity of a phonon mode can be expressed as:

\begin{equation*}
I \propto \sum_{j} {\lvert\vec{e_i}\cdot\tilde{R_j}\cdot\vec{e_a}\rvert}^2
\end{equation*}

Here, $\vec{e_i}$ and $\vec{e_a}$ are unit vectors of the incident and scattered light polarization, and $\tilde{R_j}$ is a 3$\times$3 Raman tensor. We have set the cartesian coordinates $(xyz)$ on the sample, with the laser propagating along the $[100]$ direction. Defining the sample rotation angle as $\theta$, and considering the Raman tensors of A$_g$ and B$_{3g}$ modes, which take the form based on the point group D$_{2h}$~\cite{Loudon1964}, we obtain the following expressions:

\begin{equation*}
	\tilde{R}\left(A_g\right)= \left({\begin{array}{*{20}{c}}
		a & 0 & 0 \\
		0 & b & 0 \\
		0 & 0 & c
		\end{array}} \right) \tilde{R}\left(B_{3g}\right)= \left({\begin{array}{*{20}{c}}
		0 & 0 & 0 \\
		0 & 0 & d \\
		0 & d & 0
		\end{array}} \right)
\end{equation*}

Using these tensors, we can derive the angular-dependent intensity expressions for the A$_g$ and B$_{3g}$ modes:

\begin{gather*}
I_\parallel\left(A_g\right) \propto c^2\lvert 1+\left(\frac{b}{c}-1\right)\sin^2\theta \rvert^2 \\
I_\perp\left(A_g\right) \propto \frac{1}{4}\left(b-c\right)^2 \sin^22\theta \\
I_\parallel\left(B_{3g}\right) \propto d^2 \sin^22\theta \\
I_\perp\left(B_{3g}\right) \propto d^2 \cos^22\theta
\end{gather*}

It is evident that $I_\perp\left(A_g\right)$, $I_\parallel\left(B_{3g}\right)$, and $I_\perp\left(B_{3g}\right)$ are proportional to $\sin^22\theta$ or $\cos^22\theta$. As a result, the three vibrational modes exhibit fourfold symmetry with respect to $\theta$, in agreement with our experimental results shown in Fig. 4(e)-4(g). The symmetry of $I_\parallel\left(A_g\right)$ depends on the elements of the Raman tensor, which vary for different A$_g$ modes. Therefore, it is reasonable to assume that the A$_g^5$ mode has a different symmetry than the other A$_g$ modes in the parallel-polarized configuration.

\section{Conclusion}
In summary, we have successfully exfoliated large-area mono- to few-layer Ta$_2$Pd$_3$Te$_5$, and have conducted detailed Raman spectroscopy measurements in combination with first-principles calculations. Our findings have clearly revealed all observable phonon modes, and we have extensively studied the thickness- and temperature-dependent Raman spectra. Additionally, we have observed strong anisotropy in the intensities of the Raman peaks, and have explained the symmetry of the Raman peaks using the classical Placzek approximation. Our results represent a significant advancement towards the realization of quantum devices based on few or monolayer Ta$_2$Pd$_3$Te$_5$.

\section*{Acknowledgments}
This work was supported by the Ministry of Science and Technology of China (Grant No. 2018YFE0202700), the National Natural Science Foundation of China (Grants No. 11974391, No. 11825405, No. 1192780039, and No. U2032204), the International Partnership Program of Chinese Academy of Sciences (Grant No. 112111KYSB20200012), the CAS Project for Young Scientists in Basic Research (Grant No. YSBR-047), and the Strategic Priority Research Program of Chinese Academy of Sciences (Grant No. XDB33030100).

$^\star$ These authors contributed equally to this work.

\end{document}